\documentclass[aps,twocolumn,pra,groupedaddress,superscriptaddress,nofootinbib]{revtex4}
\usepackage{graphicx,amsmath,amssymb,amsbsy,subfigure,hyperref,bbm,times,txfonts,float}
\usepackage{subfigure,hyperref,bbm,times}
\usepackage[T1]{fontenc}
\usepackage{braket}
\usepackage{epsfig}
\usepackage{color}
\usepackage{graphicx}% Include figure files
\usepackage{dcolumn}% Align table columns on decimal point
\usepackage{bm}% bold math
\newcommand{\ave}[1]{\langle#1\rangle}

\providecommand{\openone}{\leavevmode\hbox{\small1\kern-3.8pt\normalsize1}}

\usepackage{soul}

\hypersetup{
   colorlinks=true,
   linkcolor=blue,
}
\graphicspath{{figure/}}
\begin{document}

\title{Validating and controlling quantum enhancement against noise by motion of a qubit}

\author{Farzam Nosrati}
\email{farzam.nosrati@unipa.it}
\affiliation{Dipartimento di Ingegneria, Universit\`{a} di Palermo, Viale delle Scienze, Edificio 9, 90128 Palermo, Italy}
\affiliation{INRS-EMT, 1650 Boulevard Lionel-Boulet, Varennes, Qu\'{e}bec J3X 1S2, Canada}

\author{Ali Mortezapour}
\email{mortezapour@guilan.ac.ir}
\affiliation{Department of Physics, University of Guilan, P. O. Box 41335--1914, Rasht, Iran}

\author{Rosario Lo Franco}
\email{rosario.lofranco@unipa.it}
\affiliation{Dipartimento di Ingegneria, Universit\`{a} di Palermo, Viale delle Scienze, Edificio 6, 90128 Palermo, Italy}
\affiliation{Dipartimento di Fisica e Chimica, Universit\`a di Palermo, via Archirafi 36, 90123 Palermo, Italy}

\begin{abstract}
Experimental validation and control of quantum traits for an open quantum system are important for any quantum information purpose. We consider a traveling atom qubit as a quantum memory with adjustable velocity inside a leaky cavity, adopting a quantum witness as a figure of merit for quantumness assessment. We show that this model constitutes an inherent physical instance where the quantum witness does not work properly if not suitably optimized. We then supply the optimal intermediate blind measurements which make the quantum witness a faithful tester of quantum coherence. We thus find that larger velocities protect quantumness against noise, leading to lifetime extension of hybrid qubit-photon entanglement and to higher phase estimation precision. Control of qubit motion thus reveals itself as a quantum enhancer.    
\end{abstract}

\maketitle

\section{Introduction}

Quantum coherence is one of the main features of quantum systems. Theory of coherence attempts to comprehend the fundamental difference between classical and quantum worlds which leads to a better understanding of the classical-quantum boundary \cite{streltsov2017colloquium,doi:10.1142/S0217979213450197,Sperling_2015,Adesso_2016,li2012witnessing}. Also, this distinctive quantum property is considered to be the reason behind the mechanisms which ultimately lead to quantum-enhanced devices \cite{bloch2008many,ladd2010quantum,anderlini2007controlled,wang2016experimental,
barends2014superconducting,napoliPRL,castelliniPRA,winterPRL,silvaPRL,LiuPRL,Yu2019}. Several methods have been proposed to detect and quantify quantum coherence in a physical system \cite{streltsov2017colloquium}. Quantumness verification is usually performed by tomographic techniques to reconstruct the nonclassical state of the system. However, these techniques require experimental resources in terms of measurement settings which  exponentially increase with the system complexity \cite{GUHNE20091,nielsen2002quantum}. To overcome these experimental drawbacks, a quantum witness has been introduced \cite{li2012witnessing,kofler2013condition} to determine the existence of quantum coherence in the physical system. This measure helps to classify quantum or classical behaviour by direct observations in the experiment.  

A realistic quantum system interacts inevitably with its surrounding environment. Such a spontaneous interaction mainly results in destroying coherence stored in a quantum system, known as decoherence \cite{breuer2002theory}. Typically, system-environment interactions  lead to an entangled state for the system-environment ensemble. Hence, entanglement building up during the evolution is a basic mechanism underlying decoherence. In this way, manipulation and control of decoherence can lead to harnessing system-environment entanglement. It has been demonstrated that the induced steady state entanglement between an atom and its spontaneous emission excitation can be controlled by intensity, detuning and relative phase of applied fields \cite{entezar2010controllable,mortezapour2011effect,mortezapour2013phase,abazari2011phase,yang2014influence,gholipour2019quantumness}. It is noteworthy that hybrid atom-photon entanglement has found applications in new quantum tools such as quantum repeater \cite{briegel1998quantum,yuan2008experimental}, quantum networks and quantum memories \cite{rosenfeld2007remote}. 

Furthermore, it is well known that quantum coherence plays a role to achieve a more precise estimation of unknown parameters imposed by classical limitation physics. Quantum metrology allows us to reach a measurement precision that surpasses the classically achievable limit by exploiting quantum features and is becoming one of the pillars of future quantum sensors \cite{PirandolaReview}. 
In the absence of noise, the so-called Heisenberg scaling can be obtained using $N$ entangled probes in parallel \cite{huelga1997improvement, giovannetti2004quantum,giovannetti2006quantum,giovannetti2011advances,PhysRevLett.85.2733}. Quantum Fisher information (QFI), which characterizes the sensitivity of the state with respect to changes in a parameter, lies at the heart of quantum metrology \cite{giovannetti2011advances}. QFI provides a bound to distinguish the members of a family of probability distributions. For an estimation parameter with a larger QFI value, the accuracy is more clearly achieved. 
However, the decoherence can act as an external noise limiting the accuracy in the result of quantum parameter estimation which leads to loss of coherence or entanglement of the probes \cite{chin2012quantum,demkowicz2012elusive,chaves2013noisy}. It is thus important to protect the QFI from decoherence. In this regards, a substantial amount of literature has been devoted to find strategies for controlling QFI against detrimental noise \cite{berrada2013non,lu2015robust,li2015classical,berrada2015protecting,ban2015quantum,ren2016protection,liu2017quantum,huang2018protecting, mortezapour2018protecting,yang2019enhancing}.

In this work, we aim at investigating the role of qubit motion as a quantum enhancer. Firstly, the quantum witness dynamics of a moving atom inside a zero-temperature dissipative cavity is presented. This initial study is very insightful since the model naturally evidences how the quantum witness needs to be optimized for faithfully and efficiently assessing nonclassicality in an experiment. In particular, similarly to adaptive quantum tomography \cite{FlammiaNJP}, we provide the optimal quantum witness by individuating the suitable intermediate blind measurement such that it achieves its upper bound, coinciding with a coherence monotone. We then explicitly show that increasing the velocity of the atom qubit enriches the nonclassical behavior of the system. After this main result, we find that larger qubit velocities lead to extending the lifetime of hybrid entanglement between the qubit and the reservoir photon arising from atomic decay. Finally, we prove that the phase estimation precision by QFI tends to remain close to its initial maximum value thanks to qubit motion.

The paper is organized as follows. In Sec.~\ref{secBe11} we present the model, giving the explicit expression of the evolved reduced density matrix. 
Sec.~\ref{Witness} is devoted to discuss the dynamics of quantum witness and its optimization. In Sec.~\ref{AtmPht}, using von Neumann entropy, we study the time behavior of entanglement between moving qubit and cavity photon. The results concerning quantum Fisher information are presented in Sec.~\ref{Metrology}. Finally, Sec.~\ref{Conclusion} summarizes the main results.

\section{Description of the model}\label{secBe11}

\begin{figure}[t!]
   \centering
\includegraphics[width=0.5\textwidth]{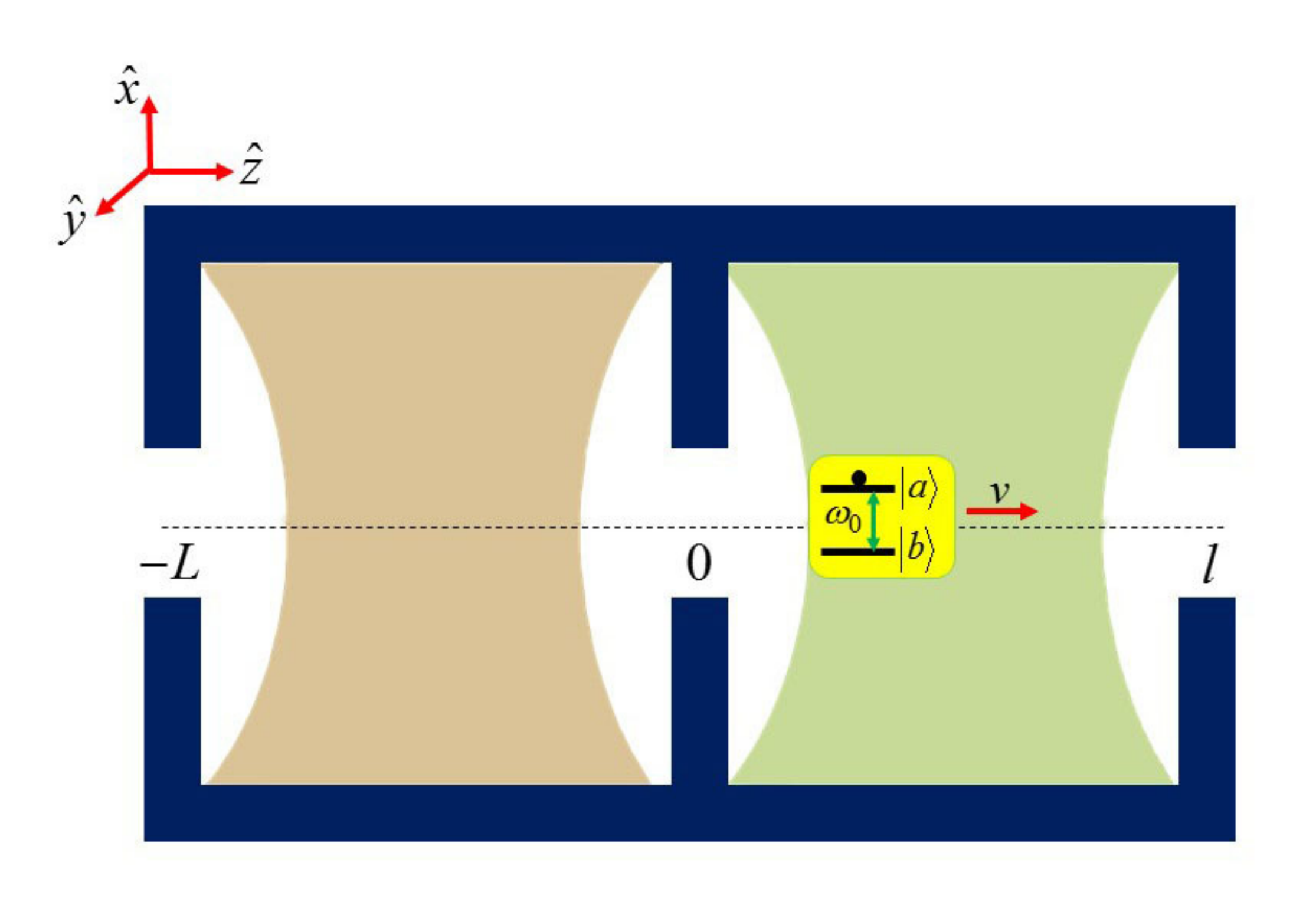}
   \caption{\label{Fig1} Schematic illustration of a setup in which a qubit (two-level atom) is traveling with constant velocity $v$ inside a cavity. The motion of the qubit is restricted along the $z$ direction (cavity axis).}
  \end{figure}  

The system under our consideration is composed of an atom qubit with transition frequency $\omega_0$ traveling inside a zero-temperature structured environment made of two perfect reflecting mirrors which are placed at $z=-L$  and $z=l$ and a partially reflecting mirror at the position $z=0$. This creates a sort of two consecutive cavities ($-L$, $0$) and ($0$, $l$), as depicted in Fig.~\ref{Fig1}. The qubit (two-level atom) is supposed to only interact with the second cavity ($0$, $l$) and moves along the $z$-axis with a constant velocity $v$. 
Such a condition can be thought to be fulfilled by Stark shifting (for instance, by turning on a suitable external electric field) the atom frequency far out of resonance from the cavity modes until $z=0$, after which the Stark shift is turned off \cite{mortezapour2017protecting}. 

During the translational motion, the qubit interacts with the cavity modes. Under the dipole and rotating-wave approximation, the Hamiltonian of the system in the interaction picture is written as ($\hbar \equiv 1$)
\begin{equation}
H_{I}= \sum _{k} f_{k}(z)[g_{k} \sigma _{+} a_{k} e^{i(\omega_{0} - \omega_{k})t} + g_{k}^{*}  a_{k}^{{\rm \dag}}{\sigma }_{-}e^{-i(\omega_{0} - \omega_{k})t}],
\label{Eq1}
\end{equation}
where $\sigma _{+}=\ket{a}\bra{b}$ ($\sigma _{-}=\ket{b}\bra{a}$ ) is the raising (lowering) operator of the qubit, with $\left| a \right\rangle$ and $\left| b \right\rangle$ respectively symbolizing the excited and ground state. In Eq.~(\ref{Eq1}), $a_{k}^{{\rm \dag }}$ ($a_{k}^{\text{}}$) denotes the creation (annihilation) operator for the $k$-th cavity mode with frequency ${{\omega }_{k}}$ and ${{g}_{k}}$ is the coupling constant between the qubit and the $k$-th mode. The parameter $f_{k} (z)$ describes the shape function of qubit motion along the $z$-axis, and it is given by \cite{leonardi1993,mortezapour2017protecting,mortezapour2017non}
\begin{equation}
\begin{aligned}
 f_{k}(z)=f_{k}(vt)=\sin[k(z-l)]=\sin[\omega_{k}(\beta t-l/c)],
\end{aligned}
\label{Eq2}
\end{equation}
where $\beta=v/c$, $c$ being the speed of light. It is evident that the coupling function will be nonzero for $z=0$ and zero for $z=l$ (perfect boundary).

It is worth mentioning that the translational motion of an atom can be considered classical ($z=vt$) as long as the de Broglie wavelength $\lambda_{B}$ of the atom is much smaller than the wavelength $\lambda_{0}$ of the resonant transition ($\lambda_{B}/\lambda_{0}\ll 1$) \cite{mortezapour2017protecting,mortezapour2017non,leonardi1993,cook1979atomic}. Moreover, the relative smallness of photon momentum ($\hbar \omega_{0}/c $) compared to atomic momentum ($mv$) allows one to neglect the atomic recoil resulting from the interaction with the electric field \cite{wilkens1992spontaneous}. These conditions can be retrieved for a $^{85}$Rb Rydberg microwave qubit ($\omega_{0}=51.1$ GHz, decay rate $\gamma=33.3$ Hz) when the velocity is $v \gg 10^{-7}$ m/s or for an optical qubit ($\omega_{0} \approx 10^{15}$ Hz, decay rate $\gamma \approx 10^{8}$ Hz) when its velocity is $v\gg 10^{-3}$ m/s \cite{PhysRevA93,RevModPhys01}. 

Let us now suppose the atomic qubit is initially prepared in the state $\ket{\psi_0}=\cos(\theta/2) \ket{a}+\sin(\theta/2)\ket{b} $ and the cavity modes in the vacuum state $\ket{0}$. As the number of excitations are conserved in this model, the total state is restricted to the single excitation manifold, admitting the closed form
\begin{equation}
\begin{aligned}
\ket{\psi(t)}= \cos(\theta/2) A(t)\ket{a}\ket{0}+\sin(\theta/2) \ket{b}\ket{0}
\\
+\sum_{k}B_{k}(t)\ket{g}\ket{{1}_k}
\end{aligned}
\label{eq:state}
\end{equation}
where $\ket{{1}_k}$ is the cavity state with a single photon in mode $k$, i.e., $\ket{{1}_k}=\hat{a}_k^{\dagger}\ket{0}$ and ${{B}_{k}}(t)$ is its probability amplitude. By substituting Eq.~(\ref{eq:state}) into the Schr\"{o}dinger equation, we obtain a dynamical equation for $A(t)$ as \cite{mortezapour2017protecting}
\begin{equation}
\begin{aligned}
\dot{A}(t)+\int_{0}^{t}{d{{t}'}}K(t,{{t}'}){A}({{t}'})=0,
\end{aligned}
\label{eq:Adiff}
\end{equation}
where the kernel $K(t,{{t}'})$, which includes the memory effects, has the form
\begin{equation}
\begin{aligned}
K(t,{{t}'})=\sum_{k}{{\vert g_{k} \vert}^2 f_{k}(vt) f_{k}(v{t}') {{e}^{-i({{\omega }_{k}}-{{\omega }_{0}})(t-{{t}'})}}},
\end{aligned}
\label{eq:kernel}
\end{equation}
This kernel expressed in the continuum limit becomes
\begin{equation}
\begin{aligned}
K(t,t' )=\int _{0}^{\infty }J(\omega _{k} )\sin [\omega _{k} (\beta t-\tau )]\sin [\omega _{k} (\beta t' -\tau )]\\
\times e^{-i(\omega _{k} -\omega _{0} )(t-t' )}  d\omega _{k} ,
\end{aligned}
\label{eq:kernel2}
\end{equation}
where $J({{\omega }_{k}})$ is the spectral density of reservoir modes. We choose a Lorentzian spectral density, which is typical of a structured cavity \cite{breuer2002theory,lofrancoreview}, whose form is
\begin{equation}
\begin{aligned}
J({{\omega }_{k}})=\frac{1}{2\pi }\frac{\gamma {{\lambda }^{2}}}{[{{({{\omega }_{0}}-{{\omega }_{k}}-\Delta )}^{2}}+{{\lambda }^{2}}]},
\end{aligned}
\label{eq:Spectrum}
\end{equation}
where $\Delta ={{\omega }_{0}}-{{\omega }_{c}}$ is the detuning between the center frequency of the cavity modes ${{\omega }_{c}}$ and ${{\omega }_{0}}$. The parameter 
$\gamma $ is related to the microscopic system-reservoir coupling constant, and $\lambda $ defines the spectral width of the coupling. It is noteworthy that the parameters $\gamma $and $\lambda $ are related to the reservoir correlation time ${{\tau }_{r}}$ and the qubit relaxation time ${{\tau }_{q}}$ as ${{\tau }_{r}}={{\lambda }^{-1}}$ and ${{\tau }_{q}}\approx {{\gamma }^{-1}}$ respectively \cite{breuer2002theory}. Qubit-cavity weak coupling occurs for $\lambda >\gamma $ (${{\tau }_{r}}<{{\tau }_{q}}$); the opposite condition $\lambda <\gamma $ (${{\tau }_{r}}>{{\tau }_{q}}$) thus identifies strong coupling. The larger the cavity quality factor, the smaller the spectral width $\lambda $.

We now recall for convenience the analytical calculation of the time-dependent coefficient $A(t)$ \cite{mortezapour2017non}. In the continuum limit ($\tau \to \infty $) and $t>t'$, analytic solution of Eq.~(\ref{eq:kernel2}) yields
\begin{equation}
\begin{aligned} 
K(t,t')=\frac{\gamma \lambda }{4} \cosh [\theta (t-t')]e^{-\bar{\lambda}(t-t')}
\end{aligned}
\label{eq:8}
\end{equation}
where $\bar{\lambda}=\lambda -i\Delta $ and $\theta =\beta (\bar{\lambda }+i\omega_{0} )$. Inserting Eq.~(\ref{eq:8}) into Eq.~(\ref{eq:Adiff}) and solving the resultant equation by Bromwich integral formula, $A(t)$ is given by
\begin{equation}
\begin{aligned} 
A(t)=\frac{(x_{1} +u_{+} )(x_{1} +u_{-} )}{(x_{1} -x_{2} )(x_{1} -x_{3} )} e^{x_{1} \gamma t} -\frac{(x_{2} +u_{+} )(x_{2} +u_{-} )}{(x_{1} -x_{2} )(x_{2} -x_{3} )} e^{x_{2} \gamma t} \\
+\frac{(x_{3} +u_{+} )(x_{3} +u_{-} )}{(x_{1} -x_{3} )(x_{2} -x_{3} )} e^{x_{3} \gamma t},
\end{aligned}
\label{eq:9}
\end{equation}
where the quantities $x_{i}$ ($i=1,2,3$) are the solutions of the cubic equation
\begin{equation}
\begin{aligned}
x^{3} +2(y_{1} -iy_{3} )x^{2} +(u_{+} u_{-} +y_{1} /4)x+y_{1} (y_{1} -iy_{3} )/4=0,
\end{aligned}
\label{eq:10}
\end{equation}
with $y_{1} =\lambda /\gamma $, $y_{2} =\omega _{0} /\gamma $, $y_{3} =\Delta /\gamma $ and $u_{\pm } =(1\pm \beta )y_{1} \pm i\beta y_{2} -i(1\pm \beta )y_{3} $.

After obtaining $A(t)$, the reduced density matrix of the qubit $\rho(t)$  can be written as
\begin{equation}
\begin{aligned}
\rho(t)=\begin{pmatrix}
  \cos^2(\theta/2)\left|{A}(t)\right| ^2 & \frac{1}{2}\sin(\theta){A}(t) \\
 \frac{1}{2}\sin(\theta){A}^*(t) & 1-\cos^2(\theta/2)\left|{A}(t)\right| ^2
 \end{pmatrix}.
 \end{aligned}
\label{eq:density}
\end{equation}
The knowledge of the evolved state of the qubit shall allow us to analyze all the physical quantities of interest to our aims. In the following, the case of resonant atom-cavity interaction ($\Delta=0$) and strong coupling ($\lambda<\gamma$) shall be considered.

\section{Quantum witness optimization}\label{Witness}

In this section, the quantum character of a moving two-level atom (qubit) in a leaky cavity is studied using a quantum witness. The general aim is to highlight the importance of the considered model to figure out the necessity to optimize the quantum witness for faithful experimental investigation of quantumness in nonisolated systems. As a main result, we shall supply the suitable  measurements to be performed on the qubit such that the quantum witness reaches its upper bound, being equal to a coherence monotone, during the evolution.

Quantum witnesses have been introduced in the literature to probe quantum coherence without resorting to demanding tomographic processes \cite{LG9,li2012witnessing,kofler2013condition}. Such witnesses reveal to be finer than the Leggett-Garg inequality \cite{LG6} and can be effectively adopted to experimentally test emergence of nonclassicality in open quantum systems. We utilize the quantum witness defined as \cite{li2012witnessing}
\begin{equation} \label{QW}  
\mathcal{W}(t)=|p_m(t)-p'_m(t)|,
\end{equation}
where $p_m(t)$ is the quantum probability to find the system in the state $m$ at time $t$, while $p'_m(t)$ represents the so-called classical probability obtained at time $t$ after an intermediate measurement has been done on the system. 
%The matrix $\Omega_\mathrm{mn}(t,t_0)$ is the propagator which permits to obtain the probability of finding the system in the state $m$ at time $t$ provided that it was in the state $n$ at time $t_0$. 
Conceptually, the quantum witness is based on the classical no-signaling theorem that an intermediate observation at time $t_0$ cannot perturb the statistical outcomes of the later measurement at time $t$, so that $p_m(t)=p'_m(t)$ ($\mathcal{W}(t)=0$) and the system behaves as a classical one \cite{li2012witnessing,kofler2013condition,munro2016}. Nonzero values of 
$\mathcal{W}(t)$ thus testify nonclassicality of the system state at time $t$. Also, the quantum witness is upper bounded by $\mathcal{W}_\mathrm{max}=1-1/d$, where $d$ is the system dimension (which equals the number of possible outcomes of a nonselective measurement on the system) \cite{schild2015maximum}.

Since the quantum and classical probabilities appearing in the quantum witness are obtained by averaging of projection operators on the system state at time $t$, it is convenient to find a propagator for the reduced density matrix of the qubit. By means of the Lindblad-type evolution for an operator X in the Heisenberg picture $d\hat{X}/dt=\mathcal{L}[\hat{X}]$ \cite{breuer2002theory,LG9}, for our dissipative system-environment model one gets the integro-differential equation 
\begin{equation}
X(t)+\int_{0}^{t}\mathrm{d}t'\mathcal{K}_t[X(t')]= 0,  
\end{equation} 
where
\begin{equation}
\mathcal{K}_t[X(t')]=K(t,t')(\sigma_{+}\sigma_{-}X(t')+X(t')\sigma_{+}\sigma_{-}-2\sigma_{+}\hat{X}(t')\sigma_{-}),
\end{equation} 
with the function $K(t,t^{\prime})$ being the kernel of Eq.~(\ref{eq:8}). Considering the evolution of the basis of Pauli operators $\{\openone, \sigma_x,\sigma_y,\sigma_z\}$, one easily obtains
\begin{gather}
   \begin{pmatrix}
  \sigma_x(t) \\\sigma_y(t) \\\sigma_z(t) \\ \openone(t)
  \end{pmatrix}= \Omega(t,0)
  \begin{pmatrix}
  \sigma_x(0) \\\sigma_y(0) \\\sigma_z(0) \\ \openone(0)
  \end{pmatrix},
  \end{gather}
where
\begin{gather}
\small
  \Omega(t,0) = \begin{pmatrix}
  \frac{1}{2}\left({A}(t)+A^*(t)\right) & \frac{-i}{2}\left({A}(t)-A^*(t)\right) & 0 & 0 \\
 \frac{i}{2}\left({A}(t)-A^*(t)\right) &  \frac{1}{2}\left({A}(t)+A^*(t)\right) & 0 & 0 \\
 0 & 0 & \left|{A}(t)\right| ^2 & \left|{A}(t)\right| ^2-1 \\
 0 & 0 & 0 & 1
  \end{pmatrix},
  \end{gather}  
with $A(t)$ given in Eq. \eqref{eq:9}. This equation can be used to directly obtain the average values of the Pauli operators at a time $t$ as $\ave{\sigma_i(t)}=\Omega(t,0)\ave{\sigma_i(0)}$ ($i=x,y,z$). Therefore, the qubit density matrix at time $t$ in the Pauli basis is given by
\begin{equation}
    \rho(t)=\frac{1}{2}\left(\openone+\ave{\sigma_x(t)}\sigma_x+\ave{\sigma_y(t)}\sigma_y+\ave{\sigma_z(t)}\sigma_z\right).
\end{equation}
The qubit, as said in Sec. \ref{secBe11}, is initially prepared in the coherent superposition 
$\ket{\psi_0}=\cos(\theta/2) |a\rangle+\sin(\theta/2)\ |b\rangle $, while the final qubit state to be measured is the maximally coherent state $\ket{+}=(\ket{a}+\ket{b})\sqrt{2}$. In the absence of intermediate measurements, the quantum probability at a time $t=\tau$ of finding the final state $\ket{+}$ is given by the expectation value of the projector $\Pi_{x,+}=\frac{1}{2}\left(\openone+\sigma_x\right)$ \cite{LG9}, that is $p_{+}(\tau)=\mathrm{Tr}(\rho(\tau)\Pi_{x,+})$ where $\rho(\tau)$ is the evolved reduced density matrix of the qubit of Eq.~(\ref{eq:density}). Differently, if a nonselective blind measurement $\{\Pi_{\pm}^\mathrm{b}\}$ is performed at intermediate time $t=\tau/2$, the state at that time becomes
\begin{equation}\label{rhoprime}
   \rho'(\tau/2)=\Pi_{+}^\mathrm{b}\rho(\tau/2)\Pi_{+}^\mathrm{b}+\Pi_{-}^\mathrm{b}\rho(\tau/2)\Pi_{-}^\mathrm{b}.
\end{equation}
By letting the perturbed state $\rho'(\tau/2)$ evolve to time $t=\tau$ leading to $\rho'(\tau)$, the classical probability is then obtained by $p'_+(\tau)=\mathrm{Tr}(\rho'(\tau)\Pi_{x,+})$. A blind measurement represents a measurement in a system basis for which the outcomes are discarded \cite{schild2015maximum}, the post-measurement state resulting in a statistical mixture corresponding to the different outcomes, as evinced by Eq.~(\ref{rhoprime}). The typical choice for the intermediate nonselective projections of Eq.~(\ref{rhoprime}) is \cite{LG9}
\begin{equation} \label{proj_b}
    \Pi_{x,\pm}^\mathrm{b}=\frac{1}{2}\left(\openone\pm\sigma_x\right).
\end{equation}
By calculating the expectation values giving the quantum and classical probabilities 
$p_+(\tau)$ and $p'_+(\tau)$, the quantum witness of Eq.~(\ref{QW}) is 
\begin{equation} \label{eq:Wq}  
\mathcal{W}(\tau)=\frac{1}{4}\left|\sin(\theta)\left(A(\tau)+A^{*}(\tau) - \frac{1}{2}(A(\tau/2)+A^{*}(\tau/2))^2 \right)\right|.
\end{equation}

It has been proven that the quantum witness of an isolated $d$-level system (qudit) exhibits a tighter upper bound given by half of the coherence monotone $\mathcal{W}(\tau)\leq C(\tau)/2\leq 1 - 1/d$ \cite{schild2015maximum,knee2018subtleties}, where $C(\tau)$ is the envelope of a quantum coherence measure of the evolved qubit density matrix. One typically employs the $l_1$-norm of coherence $C_{l_1}=\sum_{i\neq j}|\rho_{ij}|$ \cite{baumgratzPRL}. For a damped qubit in a Markovian thermal reservoir, it has been then seen that the envelope of the quantum witness, defined according to the usual intermediate and final measurements $\Pi_{\pm}^\mathrm{b}$ of Eq.~(\ref{proj_b}), indeed coincides with the coherence monotone \cite{LG9}. However, for a generic open (nonisolated) quantum system, the behavior of the quantum witness is more subtle \cite{knee2018subtleties}: it is not guaranteed that it reaches the upper bound by usual projective blind measurements and optimization procedures may be required.

\begin{figure}[t!] 
   \centering
\includegraphics[width=0.46\textwidth]{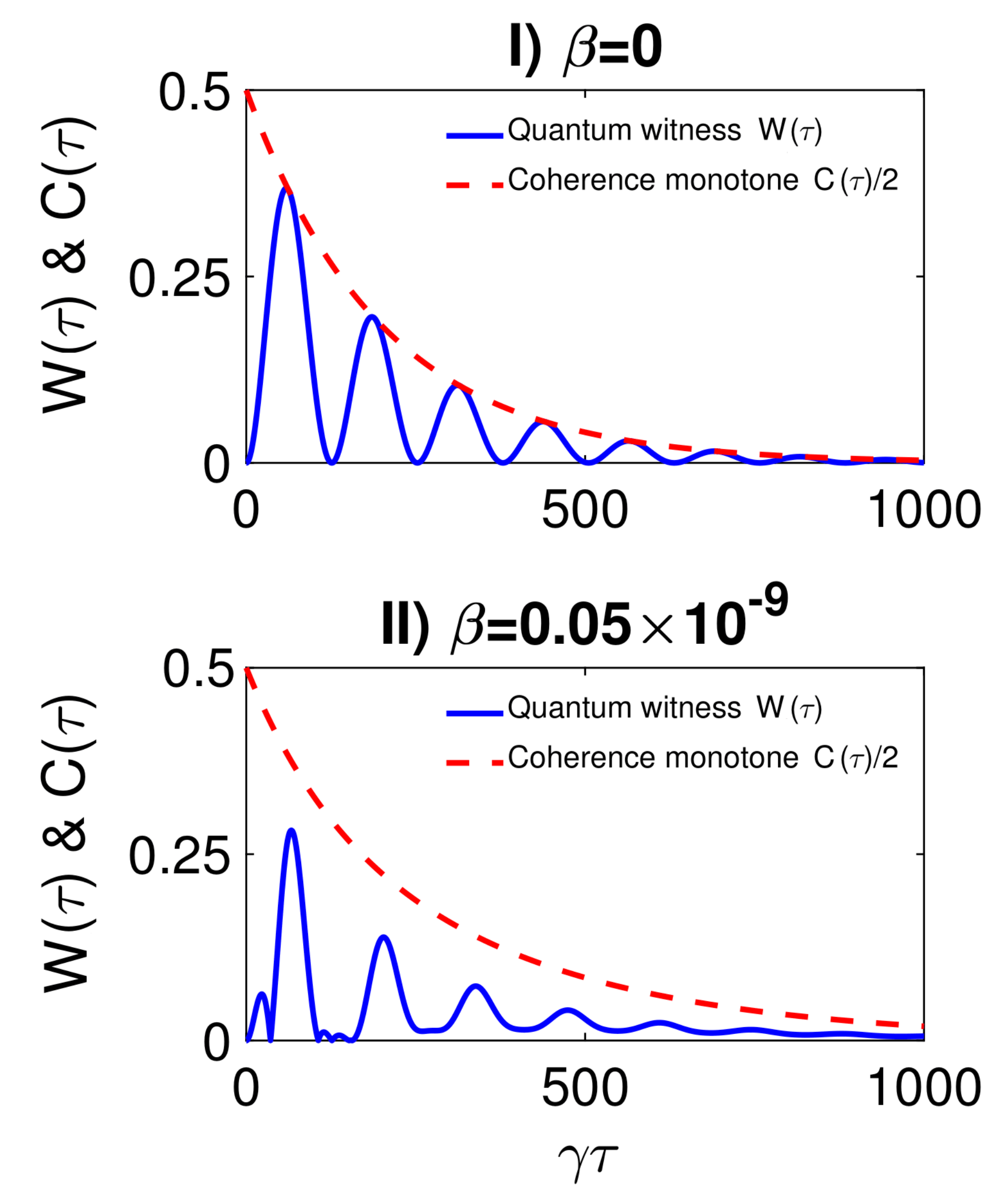}
    \caption{\label{Fig2} Quantum witness $W(\tau)$ (blue solid line) and coherence monotone $C(\tau)/2$ (red dashed line) as a function of scaled time $ \gamma \tau $ for two velocities of the qubit: I) $ \beta=0 $,  II) $ \beta=0.05\times 10^{-9}$. Others parameters are taken as: $ \theta=\pi/2 $, $ \lambda=0.01\gamma $, $ \Delta=0 $ and $ \omega_{0}=51.1 \times 10^{9} $ Hz (these values are of experimental reaching \cite{kuhr}.)}
\end{figure}

We now show how this aspect of experimental interest naturally emerges in our model. From the qubit reduced density matrix of Eq.~(\ref{eq:density}), we straightforwardly obtain $C_{l_1}(t)=|\sin(\theta) A(t)|$ for the $l_1$-norm of coherence. The quantum witness $\mathcal{W}(\tau)$ of Eq.~(\ref{eq:Wq}) and the coherence monotone $C(\tau)/2$ are then plotted as functions of dimensionless time $\gamma\tau$ for various qubit velocities. In the case of stationary qubit ($\beta=0$), as displayed in Fig. \ref{Fig2}(a), the quantum witness amplitude reaches its maximum violation coinciding with the coherence monotone, in accordance with previous results \cite{schild2015maximum, friedenberger2017assessing,friedenberger2018tailoring}. On the other hand, 
Fig. \ref{Fig2}(b) shows that, by increasing the velocity of qubit $\beta$, the quantum witness maximum values decrease in spite of an increase of the coherence monotone curve. Therefore, we infer that the intermediate nonselective projections of Eq.~(\ref{proj_b}) are not optimally selected for a general open system dynamics. 
Some quantum coherence witnesses for nonisolated systems have been constructed \cite{knee2018subtleties} by using, as intermediate perturbation, the so-called classicalization operation $\Gamma(\rho(\tau/2))\equiv\sum_{i}\ket{i}\bra{i}\rho(\tau/2)\ket{i}\bra{i}$, which is the formal process that preserves the diagonal entries of the system state but destroys the off-diagonal ones. Such a classicalization can be experimentally simulated by randomization of the phase of path-encoded photonic qudits \cite{wang2017optimal}. 

%Although a measurement of the preferred basis is performed (the populations of the various classical states are inferred), it is not through a process which can be modeled by projective measurement operators. The solution that they concentrate on in Ref. \cite{knee2018subtleties} is to infer the full set of probabilities and then reprepare the appropriately weighted mixture of classical states from a fiducial state, resulting in $\Gamma(\rho)$.

\begin{figure}[t!] 
   \centering
\includegraphics[width=0.46\textwidth]{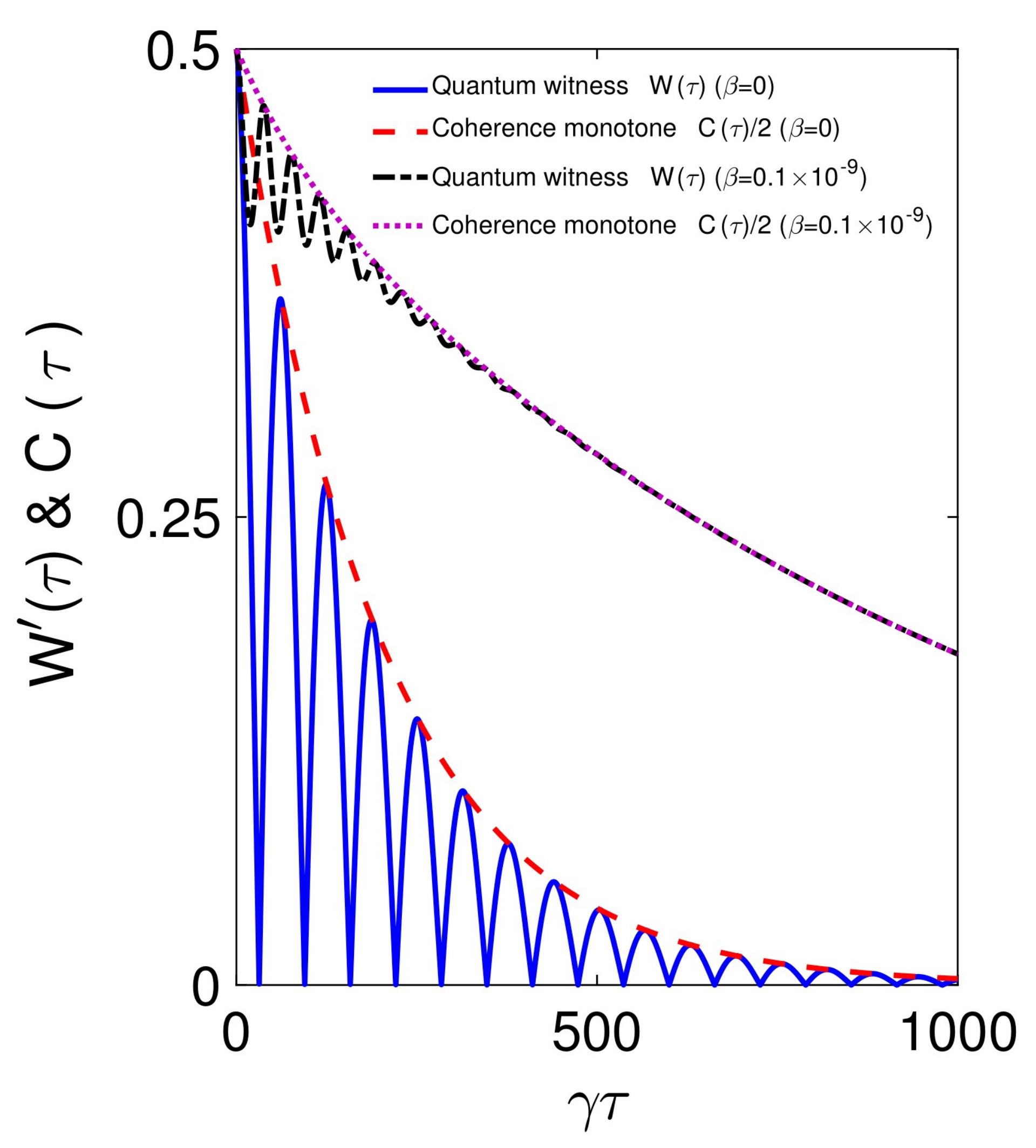}
    \caption{\label{Fig3} Optimized quantum witness $ \mathcal{W}'_\mathrm{q}(\tau)$ and coherence monotone $C(\tau)/2$ as a function of scaled time $\gamma \tau $ for two different velocities of the qubit, $\beta=0$ (bottom curves) and $ \beta=0.1\times 10^{-9}$ (top curves). Others parameters are the same as those of Fig.~\ref{Fig2}.}
\end{figure}

Inspired by these arguments, we want to provide here a simple experimentally-feasible method to optimize the quantum witness of Eq.~(\ref{QW}). As a matter of fact, one needs suitable blind intermediate measurements which make the system state classical (incoherent) so that it can remain classical for the remainder of the evolution. We remark that, once such measurements are found, they work for any open system dynamics arising from an incoherent channel, that is a channel incapable of creating quantum coherence in the state of the system \cite{DATTA2018243,jianweiPRA}. Since one is interested in making the system state classical in the preferred computational basis, we find that the goal is inherently-accomplished by the nonselective projections 
\begin{equation}\label{Piz}
    \Pi_{z,\pm}^\mathrm{b}=\frac{1}{2}\left(\openone\pm\sigma_z\right),
\end{equation}
which have to be substituted in Eq.~(\ref{rhoprime}) to give the new intermediate mixed qubit state $\rho'(\tau/2)$, that results to be diagonal (classical).
Performing these new blind measurements and letting the perturbed state $\rho'(\tau/2)$ evolve to time $\tau$, the qubit state at time $t=\tau$ is
\begin{equation}
\begin{aligned}
\rho'(\tau)=\begin{pmatrix}
  \cos^2(\theta/2)\left|{A}(\tau)\right| ^4 & 0 \\
 0 & 1-\cos^2(\theta/2)\left|{A}(\tau)\right| ^4
 \end{pmatrix},
 \end{aligned}
 \label{eq:density2}
\end{equation}
which remains, as desired, a classical mixture (the dissipative channel of our model is incoherent for the qubit). 
Notice that the diagonal elements of $\rho'(\tau)$ are different from the diagonal elements of $\rho(t)$ of Eq.~(\ref{eq:density}).
Calculating the quantum and classical probabilities with the usual final measurement defined by the projector $\Pi_{x,+}$ (final measured qubit state 
$\ket{+}$), we obtain
\begin{equation} \label{eq:Wqprime}  
\mathcal{W}'(\tau)=\frac{1}{4}\left| \sin(\theta)\left(A(\tau)+A^{*}(\tau)\right)\right|=\frac{1}{2}|\sin(\theta) \mathcal{R}(A(\tau))|.
\end{equation}
The quantum witness $\mathcal{W}'(\tau)$ is now optimal and, as deduced from Eq.~(\ref{eq:density}), it coincides with the real part of the off-diagonal term of the evolved qubit density matrix. This optimization procedure is thus clearly due to an adaptive blind measurement, in analogy with adaptive quantum tomography \cite{FlammiaNJP}. This result can be also interpreted as maximizing the distance between the state of the system at time $t=\tau$ and its perturbed counterpart, which results to be a classical state. As displayed in Fig.~\ref{Fig3}, $\mathcal{W}'(\tau)$ now reaches its upper bound (coherence monotone) during the dynamics for any value of qubit velocity, which guarantees a faithful use of the quantum witness with adaptive blind measurements as an experimentally-friendly coherence tester. As an immediate byproduct of this fact, Fig.~\ref{Fig3} shows that the preservation of quantum witness is extended by increasing the velocity of the qubit. In other words, the motion of the qubit acts as a shield to protect quantum memory stored in the qubit against noise. 

\section{Hybrid qubit-photon entanglement protection}\label{AtmPht}

After supplying our main result in the previous section, we would like to analyze the consequences of qubit motion on the dynamics of other useful quantum features of the system. Being in the presence of a qubit-environment interaction at zero temperature, a typical trait of interest for a comprehensive characterization of the overall system, strictly linked to the decoherence process \cite{breuer2002theory,Schloss2019}, is the formation of hybrid qubit-photon entanglement. Besides this aspect, dealing with the entanglement between quantum memory stored in a qubit and reservoir photon is relevant for implementing distribution of quantum states in quantum networks \cite{hackerNatPhot,leentArxiv,dehghaniJOSA,sabeghSciRep,Beaudoin_2017,zollerPRL}. In this section, we thus investigate the way atomic qubit velocity affects the dynamics of the entanglement established between the qubit itself and the photon due to atom excitation decay. 

A number of useful measures are available to quantify entanglement of general composite systems \cite{horodecki2009quantum}. Among these, the von Neumann entropy of a reduced density matrix quantifies the entanglement between subsystems of a composite system in a pure state. For a given state $\rho$, its von Neumann entropy is defined as $S(\rho)=-\mathrm{Tr}(\rho \ln\rho)$. In this context, it is useful to recall that, for a bipartite quantum system, the entropies of the overall system and of the subsystems satisfy the inequalities \cite{araki2002entropy}
\begin{equation}
\label{eq:Araki}
\vert S(\rho_{A}(t))- S(\rho_{F}(t))\vert  \leq S(\rho_{AF}(t)) \leq S(\rho_{A}(t))+ S(\rho_{F}(t)),
\end{equation}
where the subscripts $A$ and $F$ refer to two generic subsystems which, in our case, are atom and radiation field, respectively, while $\rho_{AF}(t)$ denotes the density matrix of the global atom-field system. 
Being the overall evolution unitary, if the global atom-field system is initially prepared in a pure state, Eq.~(\ref{eq:Araki}) conveys that the entropies of atom and field will be equal during the entire evolution: $ S(\rho_{A}(t)) =S(\rho_{F}(t)) $ \cite{phoenix1991establishment}. Under this circumstance, von Neumann entropy of a subsystem reduced density matrix actually identifies entanglement between the subsystems. Therefore, an increasing entropy tells us that the two subsystems tend to get entangled, while a decreasing entropy discloses that each subsystem evolves towards a pure quantum state and becomes unentangled \cite{phoenix1991establishment}.

\begin{figure}[t!]
   \centering
\includegraphics[width=0.45\textwidth]{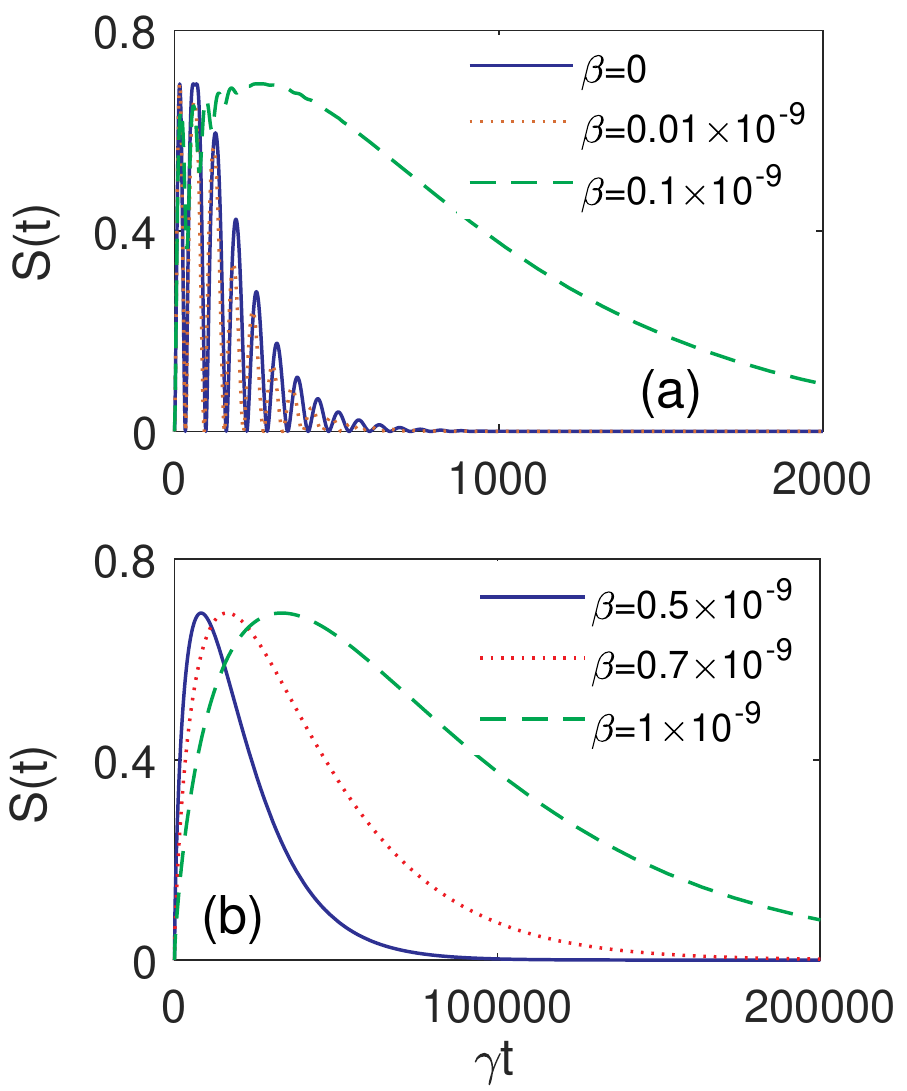}
   \caption{\label{Fig4}The von Neumann entropy of the qubit $S(t)\equiv S(\rho(t))$ as a function of $ \gamma t $ for various velocities of the qubit: (a) $ \beta=0 $ (solid-blue line), $ \beta=0.01 \times 10^{-9} $ (dotted-red line), $ \beta=0.1 \times 10^{-9} $ (dashed-green line) and (b) $ \beta=0.5 \times 10^{-9} $ (solid-blue line), $ \beta=0.7 \times 10^{-9} $ (dotted-red line), $ \beta=1 \times 10^{-9} $ (dashed-green line). Others parameters are: $ \lambda=0.01\gamma $, $ \Delta=0 $, $ \theta=\pi/2 $ and $ \omega_{0}=51.1 \times 10^{9} $ Hz.}
  \end{figure}

In Fig.~\ref{Fig4}, the von Neumann entropy $S(\rho(t))$ of the reduced state of the qubit, given in Eq.~(\ref{eq:density}), is plotted for various velocities of the qubit as a function of the scaled time $\gamma t$, starting with the qubit in the maximally coherent state $\ket{+}= (\left| a \right\rangle + \left| b \right\rangle)/\sqrt{2}$ ($\theta=\pi/2$). In the case of stationary qubit ($\beta = 0$), one observes that the qubit and its radiation field photon become entangled immediately after the qubit-cavity interaction is switched on, as expected, this entanglement eventually damping with an oscillatory behavior. On the other hand, as is manifest from Fig.~\ref{Fig4}(a-b), increasing the velocity of the traveling qubit not only remarkably lengthens the lifetime of the hybrid entanglement but it also suppresses the fluctuations (due to memory effects). From a quantitative viewpoint, one can notice that increasing the velocity of an order of magnitude results in prolonging the qubit-photon entanglement lifetime of two orders of magnitude. 
 
\begin{figure}[t!]
   \centering
\includegraphics[width=0.45\textwidth]{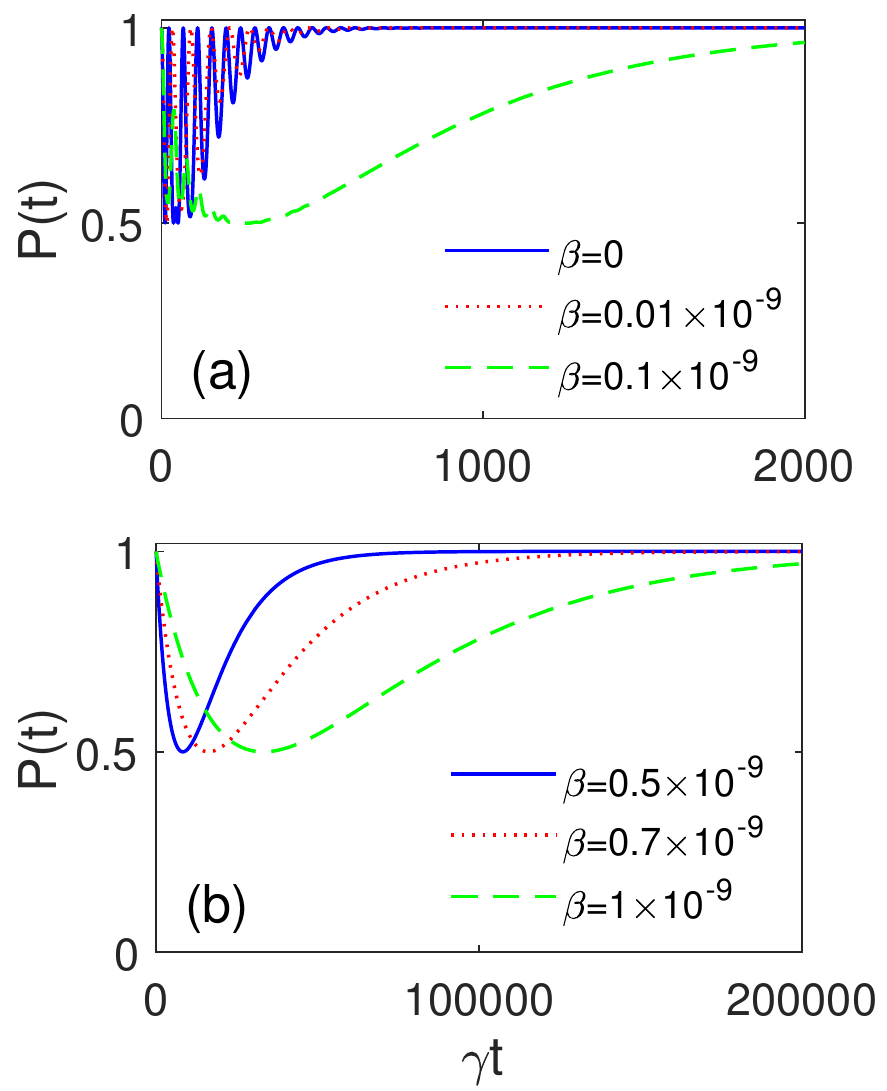}
   \caption{\label{Fig5} Purity of the qubit $ P(t) $ as a function of scaled time $ \gamma t $ for various velocities of the qubit. The other parameters are the same as those used in Fig. \ref{Fig2}.}
  \end{figure}

It is worth to highlight the relationship between entanglement entropy and qubit purity $P(t)=\mathrm{Tr}[\rho(t)^2]$. From Eq.~(\ref{eq:density}), the time-dependent purity of the qubit is
\begin{equation}
\begin{aligned}
P(t)=2\cos^4(\theta/2){\vert A(t) \vert}^2[{\vert A(t) \vert}^2-1]+1,
 \end{aligned}
\end{equation} 
whose evolution corresponding to the various velocities of the qubit is reported in 
Fig.~\ref{Fig5}. As the curves clearly illustrate, a faster qubit motion delays the reaching of the final pure ground state $\left| b \right\rangle$ for the qubit. In fact, the plots of Fig.~\ref{Fig5} and Fig.~\ref{Fig4} certify a close connection between the hybrid qubit-photon entanglement and purity of the qubit. The time during which the qubit state remains mixed is longer for larger velocities, as well as the entanglement lifetime is extended. Indeed, we can observe that the time behaviors of qubit-photon entanglement and qubit purity are symmetrical, following the same timescales.  

\section{Improving quantum phase estimation precision}\label{Metrology}

To complete our analysis, in this section we shall understand to which extent the qubit motion can affect a quantum metrology task. Quantum metrology utilizes quantum-mechanical features to improve the precision of measurements. Tipically, the parameter $\phi$ to be estimated is encoded on a probe state $\rho_\mathrm{in}$ by means of a unitary process $U_{\phi}$, giving the output state $\rho_{\phi}= U_{\phi} \rho_\mathrm{in} U_{\phi}^{\dagger}$. This output state $\rho_{\phi}$ is successively measured by means of a set of positive operator valued measurements and the value of $\phi$ finally estimated from the outcomes \cite{giovannetti2011advances}.

The so-called quantum Fisher information (QFI) is used as a criterion to quantify how precise the parameter measurement and is defined as \cite{helstrom1976quantum}
\begin{equation} 
\label{eq:15}
F_{\phi}=\mathrm{Tr}[\rho_{\phi}L^2],
\end{equation} 
where $ \rho_{\phi} $ is the density matrix of the system and $ \phi $ is the parameter to be estimated. Moreover, the symmetric logarithmic derivative operator $ L $ is meant to be a Hermitian operator fulfilling the condition $ \partial_{\phi}\rho_{\phi} =\partial \rho/\partial\phi=\lbrace L , \rho_{\phi} \rbrace $,  with $ \lbrace \cdot , \cdot \rbrace $ indicating the anticommutator \cite{helstrom1976quantum,probabilistic1982statistical}. An essential feature of the QFI is to mark a lower bound of uncertainty in parameter estimation, defined by the quantum Cramer-Rao inequality as \cite{helstrom1976quantum,probabilistic1982statistical}
\begin{equation} 
\label{eq:16}
\delta \phi \geq \delta\phi_\mathrm{min} = 1/\sqrt{F_{\phi}},
\end{equation} 
where $ (\delta \phi)^2 $ is the mean square error in the measure of parameter $ \phi $. The above inequality determines the smallest possible uncertainty in estimation of the parameter of interest. By diagonalizing the matrix $ \rho_{\phi} $ as $ \rho_{\phi}=\sum_{m} p_{m} \left| \psi_m \right\rangle \left\langle \psi_{m}\right| $, where $ p_m $ and $ \ket{\psi_m} $ are, respectively, eigenvalues and eigenstates, one can rewrite the QFI as \cite{zhang2013quantum}
\begin{equation} \label{eq:17}
F_{\phi}=\sum _{m,n} \frac{2}{p_m + p_n}
|\bra{p_m}\partial_{\phi}\rho_{\phi}\ket{p_n}|^2.
\end{equation} 
For an open quantum system, because of the inevitable detrimental role of surroundings, the quantum enhancement for parameter estimation is hindered and tends to be lose when time goes by \cite{giovannetti2011advances}. This means that the time-dependent QFI 
$F_{\phi}(t)$ is susceptible to decrease during the system evolution. So, it is important to devise techniques and strategies which can prevent this drawback. 

\begin{figure}[t!]
   \centering
\includegraphics[width=0.47\textwidth]{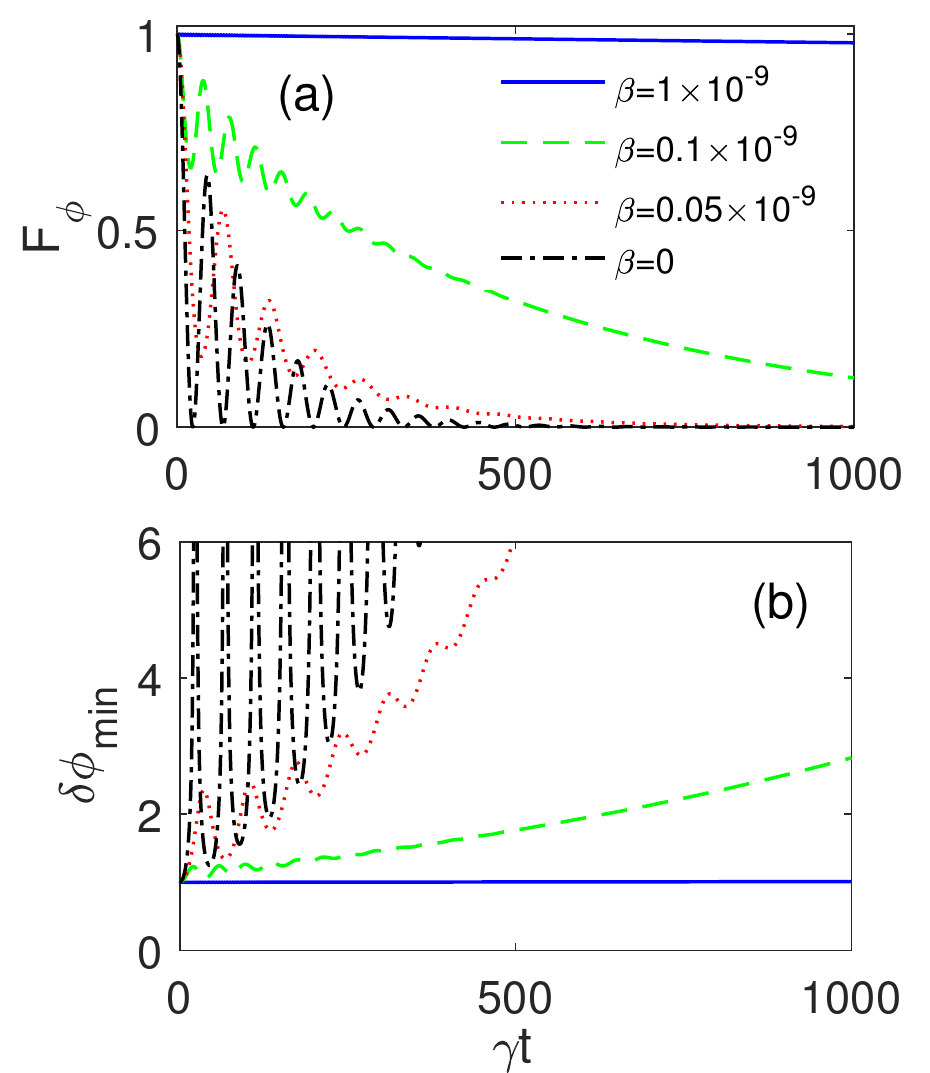}
   \caption{\label{Fig6} (a) Quantum Fisher information $ F_{\phi}(t) $ and optimal phase estimation $\delta\phi_\mathrm{min}(t)$ as a function of $\gamma t$ for various velocities of the qubit:  $\beta=0$ (dash-dotted black line), $ \beta=0.05 \times 10^{-9} $ (dotted-red line), $\beta=0.1 \times 10^{-9}$ (dashed-green line) and  $\beta=1 \times 10^{-9}$ (solid-blue line). Others parameters are taken as: $\lambda=0.01\gamma$, $\Delta=0$, $\theta=\pi/2$ and $\omega_{0}=51.1 \times 10^{9}$ Hz.}
  \end{figure}

To show the effects of qubit motion within this context, we focus on phase estimation. In particular, the (black-box) unitary $ U_\phi = \ket{b}\bra{b} + e^{i\phi}\ket{a}\bra{a} $ acts on the initial maximally coherent state of the atomic qubit 
$ \ket{+}=(\ket{a}+\ket{b})/\sqrt{2} $, which is successively subjected to the open dynamics due to the interaction with the leaky cavity described in Sec.~\ref{secBe11}. The initial overall atom-cavity state is therefore $ (U_\phi \ket{+})\ket{0} $ and the evolved reduced density matrix of the qubit $\rho_{\phi}(t)$ has the same form of Eq.~(\ref{eq:density}) where the off-diagonal elements now depend on the phase $\phi$. 
In Fig.~\ref{Fig6} the time evolution of $F_{\phi}(t)$ and of the optimal phase estimation $\delta\phi_\mathrm{min}(t)$ is displayed for different velocities of the qubit. From 
Fig.~\ref{Fig6}(a-b) one can observe that, compared to the case of stationary qubit ($\beta = 0$), larger velocities significantly inhibit the decrease of QFI and, as a consequence, maintain the uncertainty $\delta\phi_\mathrm{min}(t)$ close to its initial value. Quantum-enhancement for phase estimation is thus maintained thanks to the qubit motion, despite the dissipative noise, with the further advantage of stabilizing the error by quenching the oscillations (due to the memory effects). 

\section{Conclusion}\label{Conclusion}

In this work we have investigated the role of qubit motion in the maintenance against noise of the quantum character of the qubit, which is assessed by the directly-measurable quantum witness. The qubit has been chosen as a two-level atom traveling inside a dissipative high-$Q$ cavity at zero temperature (see Sec.~\ref{secBe11}). The choice of this model has proven to be suitable for enlightening the problem of optimizing the quantum witness in an open system dynamics. The quantum witness definition depends on blind measurements to perform on the system of interest at an intermediate time of the evolution. A faithful experimental use of the quantum witness as a measure of quantum coherence assumes that the measurements are such that it can reach its upper bound. We have shown that, using the typical measurements projecting the qubit onto a maximally coherent state in the computational basis (see Sec.~\ref{Witness}), the quantum witness decreases in spite of a coherence gain for nonzero qubit velocity. We point out that a physical instance where this mismatch between quantum witness and coherence clearly emerged has remained elusive so far. We have then provided the optimal intermediate blind measurements which make the quantum witness reach its upper bound during the evolution, coinciding with a coherence monotone, independently of qubit velocity. Such blind measurements are those causing the perturbed intermediate state to become a classical one ($\Pi_{z,\pm}^\mathrm{b}$, see Eq.~(\ref{Piz})), so that any incoherent channel maintains it classical for the subsequent evolution. This optimization result for the quantum witness takes on experimental interest and can be straightforwardly generalized to a system of $N$ noninteracting qubits individually coupled to their own reservoir. In fact, in this case the set of intermediate single-qubit blind measurements 
\begin{equation}
    \Pi^{\mathrm{b},(N)}_{z,\pm}=\Pi^{\mathrm{b},1}_{z,\pm}\otimes\Pi^{\mathrm{b},2}_{z,\pm}\otimes\dots\otimes \Pi^{\mathrm{b},N}_{z,\pm},
\end{equation}
makes the $N$-qubit state classical (diagonal) in the computational basis, which remains classical provided that each noisy channel is incoherent \cite{DATTA2018243,jianweiPRA}.

As a byproduct of the above main result, we have found the general behavior that larger velocities of the qubit strongly protect quantumness against noise. In particular, we have seen that this fact leads to lifetime extension of hybrid entanglement between the atom qubit and the reservoir photon arising from atomic decay (Sec.~\ref{AtmPht}). Moreover, we have proven that phase estimation precision is significantly improved and stabilized, despite the environmental noise, with the quantum Fisher information remaining closer and closer to its initial value for higher velocities (Sec.~\ref{Metrology}). 

We remark that the parameters used in this work are realistic and typically encountered in cavity-QED and circuit-QED experiments. For instance, ultrahigh finesse Fabry-Perot superconducting cavities with quality factors $Q\geq 10^{10}$, corresponding to spectral width $\lambda \leq 7$ Hz ($\tau_r \geq 130$ ms) at central frequency $\omega_c =\omega_0 \approx 51.1$ GHz, have been built \cite{kuhr,assematPRL}. In addition, high-quality cavities and controlled qubit-environment interactions can be nowadays implemented by circuit-QED technologies \cite{PaikPRL,mottonenNPJ}. Interestingly, position-dependent qubit-cavity coupling strength, described by a sinusoidal function analogous to that of Eq.~(\ref{Eq2}), can be produced in circuit-QED \cite{jonesSciRep,shanksNatComm}: the model of moving qubit here considered may be thus realized by adjusting the position of the qubit linearly with time, so to have a relation like $z = vt$.

The results of this work, besides supplying a reliable method to optimize the quantum witness, demonstrate that control of the qubit motion acts as a quantum enhancer and supply further insights towards shielding quantum features against noise.

%\bibliography{references}

\end{document}